\documentclass[a4paper]{jpconf}
\usepackage{graphicx}

\newcommand{\bq}    {\begin{equation}}
\newcommand{\eq}    {\end{equation}}
\newcommand{\bqr} {\begin{eqnarray}}
\newcommand{\eqr} {\end{eqnarray}}

\begin{document}
\title{Magnetic fields generated by r-modes in accreting millisecond pulsars}

\author{Carmine Cuofano and Alessandro Drago}

\address{Dipartimento di Fisica, Universit\'a di Ferrara \\
and INFN sez. Ferrara, 44100 Ferrara, Italy}
\ead{cuofano@fe.infn.it}

\begin{abstract}
In millisecond pulsars the existence of the Coriolis force allows the 
development of the so-called Rossby oscillations (r-modes) which are know to 
be unstable to emission of gravitational waves. These instabilities are 
mainly damped by the viscosity of the star or by the existence of a strong 
magnetic field. A fraction of the observed millisecond pulsars are known to 
be inside Low Mass X-ray Binaries (LMXBs), systems in which a neutron star 
(or a black hole) is accreting from a donor whose mass is smaller than 1 
$M_\odot$. Here we show that the r-mode instabilities can generate strong toroidal 
magnetic fields by inducing differential rotation.
In this way we also provide an alternative scenario
for the origin of the magnetars. 
\end{abstract}

\section{Introduction}

The r-mode oscillations in all rotating stars are unstable for emission of 
gravitational waves \cite{Andersson:2000mf}. These modes play therefore a 
very important role in the astrophysics of compact stars and in the search 
for gravitational waves. On the other hand the existence of millisecond pulsars 
implies the presence of damping mechanisms of the r-modes. Damping mechanisms 
are associated with bulk and shear viscosity and with the possible existence 
of the so called Ekman layer. The latter is located at the interface between 
the solid crust and the fluid of the inner core and in this region friction 
is significantely enhanced respect to friction in a purely fluid component. 
All these mechanisms are strongly temperature dependent. \\
An important class of rapidly rotating neutron stars are the accreting 
millisecond pulsars associated with Low Mass X-ray Binaries (LMXBs). For 
these objects the internal temperature is estimated to be in the range 
$10^{8}$-$10^{8.5}$ K \cite{Brown:1997ji,Bhattacharyya:2001sw} and their
frequencies can be as large as $\sim 650$ Hz. 
In this range of temperatures and in the case of a purely nucleonic star, 
bulk and shear viscosities alone cannot stabilize stars whose frequency 
exceeds $\sim 100$ Hz. A possible explanation of the stability of stars 
rotating at higher frequencies is based on the Ekman layer, but recent 
calculations show that this explanation holds only for rather 
extreme values of the parameters \cite{Bondarescu:2007jw}. In this contribution 
we propose a new damping mechanism based on the generation inside the star of 
strong magnetic fields produced by r-mode instabilities. This same mechanism has been 
proposed in the case of rapidly rotating, isolated and newly born neutron 
stars in \cite{Rezzolla:2001di,Rezzolla:2001dh}. In that paper the mechanism 
which generates the magnetic field is investigated only during the relatively short period 
in which the star remains always in the instability region. In our work
we consider accreting stars and we investigate the interplay between r-modes
and magnetic field on an extremely long period and we show that in this
scenario a very strong magnetic field can be produced. 

\section{R-mode equations in the presence of magnetic field}

R-mode instabilities are associated to kinematical secular effects 
wich generate differential rotation in the star and large scale mass drifts, 
particularly in the azimuthal direction. Differential rotation in turn
can produce very strong toroidal magnetic fields in the nucleus and these 
fields damp the instabilities extracting angular momentum from the modes.
In order to derive the equations regulating the evolution of r-modes in the
presence of a pre-existent poloidal magnetic field we have modified the 
equations derived in \cite{Wagoner:2002vr}, taking into account also the 
magnetic damping. \\
We use the estimate given in \cite{Andersson:2000mf} for the gravitational 
radiation reaction rate due to the $l = m = 2$ current multipole
\begin{equation}
F_{g} = \frac{1}{47}M_{1.4}R_{10}^{4}P^{-6}_{-3} \,\,\,\,\, \mbox{s}
\label{eq0A}
\end{equation}
as well as the bulk and shear viscosity damping rates
\begin{eqnarray}
F_{b} = \frac{1}{2.7\times 10^{11}} M^{-1}_{1.4}R_{10}
P^{-2}_{-3}T^{6}_{9} \,\,\,\,\, \mbox{s} \nonumber \\
F_{s} = \frac{1}{6.7\times 10^{7}}
M^{5/4}_{1.4}R_{10}^{-23/4}T^{-2}_{9} \,\,\,\,\, \mbox{s}
\label{eq0B}
\end{eqnarray}
where we have used the notation $M_{1.4}=M/1.4 M_{\odot}$, $R_{10}=R/10$ Km,
$P_{-3}=P/1$ ms and $T_9=T/10^9$ K.
\\
The total angular momentum $J$ of a star can be decomposed into a equilibrium
angular momentum $J_{*}$ and a perturbation proportional to the canonical
angular momentum of a r-mode $J_{c}$:
\begin{equation}
J=J_*(M,\Omega)+(1-K_j)J_c, \,\,\,\,\,\,\,\,\,\,\,\,\,  J_c=-K_c\alpha^2J_*
\label{eq1}
\end{equation}
where $K_{(j,c)}$ are dimensionless constants and
$J_{*}\cong I_{*}\Omega$. \\
Following Ref.\cite{Friedman:1978hf} the canonical angular momentum
obeys the following equation:
\begin{equation}
dJ_c/dt =2J_c\{F_g(M,\Omega)-[F_v(M,\Omega,T_v)+F_{m_i}(M,\Omega,B_p)]\}
\label{eq2}
\end{equation}
where $F_v=F_s+F_b$ is the 
viscous damping rate and we have introduced the magnetic damping rate 
$F_{m_i}$ that we discuss in the next section. \\
The total angular momentum satisfies 
instead the equation:
\begin{equation}
dJ/dt=2J_c F_g+\dot{J}_a(t)-I_{*}\Omega F_{m_e}
\label{eq3}
\end{equation}
where $\dot{J}_a$ is the rate of accretion of angular momentum and
we have assumed it to be $\dot{J}_a=\dot{M}(GMR)^{1/2}$, and $F_{m_e}$ is 
the magnetic braking rate associated to the poloidal magnetic field. \\
Combining the equations (\ref{eq2}) and (\ref{eq3}) than we give the dynamical 
evolution  relation
\begin{eqnarray}
\frac{d\alpha}{dt} &=& \alpha(F_g-F_v-F_{m_i})
+\alpha[K_jF_g+(1-K_j)(F_v+F_{m_i})]K_c\alpha^2
-\frac{\alpha\dot{M}}{2\tilde{I}\Omega}
\left(\frac{G}{MR^3}\right)^{1/2}+\frac{\alpha F_{m_e}}{2} \nonumber \\
\frac{d\Omega}{dt} &=& -2K_c\Omega\alpha^{2}
[K_jF_g+(1-K_j)(F_v+F_{m_i})]-\frac{\dot{M}\Omega}{M}
+\frac{\dot{M}}{\tilde{I}}\left(\frac{G}{MR^3}\right)^{1/2}
-\Omega F_{m_e}
\label{eq4}
\end{eqnarray}
where $I=\tilde{I}MR^2$ with $\tilde{I}=0.261$ for an n=1 polytrope.
The previous equations can be simplified when the star is close to
the instability region identified by the condition 
$F_g-F_v-F_m \approx 0$ and we obtain:
\begin{eqnarray}
\frac{d\Omega}{dt} &=& -2\Omega\alpha^2K_c\left(F_v+F_m\right)
-\frac{\dot{M}}{M}\Omega+\frac{\dot{M}}{\tilde{I}}\
\left(\frac{G}{MR^3}\right)^{1/2} \nonumber \\
\frac{d\alpha}{dt} &=& \alpha F_g-\alpha\left(F_v+F_m\right)
(1-\alpha^2K_c)-\frac{\alpha\dot{M}}{2\tilde{I}\Omega}
\left(\frac{G}{MR^3}\right)^{1/2} 
\label{eq5}
\end{eqnarray}
We note that $K_c=9.4\times 10^{-2}$ and that in this approximation the value of $K_j$ is unimportant. Moreover the magnetic breaking is negligible for magnetic fields typical of the LMBXs ($10^8$-$10^9$G) and for accretion rates considered ($10^{-8}$-$10^{-9}$ $\mbox{M}_{\odot} \mbox{yr}^{-1}$).   \\
Viscosity depends critically on temperature. We include three factors in modelling the
temperature evolution: modified URCA cooling, shear viscosity reheating, and accretion heating.
The cooling rate due to the modified URCA reactions, $\dot{\epsilon}_{u}$, is given 
in \cite{Shapiro83}
\begin{equation}
 \dot{\epsilon}_{u}=7.5\times 10^{39}M_{1.4}^{2/3}T_{9}^{8} \,\,\,\,\,\, \mbox{erg s}^{-1}
\end{equation}
The neutron star will be heated by the action of shear viscosity on the
r-mode oscillations. The heating rate due to shear viscosity, $\dot{\epsilon}_{s}$,
is given by \cite{Andersson:2000mf}
\begin{equation}
 \dot{\epsilon}_s = 2\alpha^2\Omega^2MR^2\tilde{J}F_s 
 =8.3\times10^{37}\alpha^2\Omega^2\tilde{J}M_{1.4}^{9/4}
 R_{10}^{-15/4}T_{9}^{-2} \,\,\,\,\,\, \mbox{erg s}^{-1}
\end{equation}
where $\tilde{J}=1.635\times 10^{-2}$. \\
Accretion heating have two components. We use the estimates given in \cite{Watts:2001ej}.
The first contribution arises when accreting matter undergoes nuclear 
burning at the surface of the star
\begin{equation}
 \dot{\epsilon}_n=\frac{\dot{M}}{m_{B}}\times 1.5 \mbox{MeV} = 4\times 10^{51} 
 \dot{M}_{1.4} \,\,\,\,\, \mbox{erg s}^{-1}
\end{equation}
where $m_B$ is the mass of a barion. \\
The second contribution arise becouse the flow is assumed to be
advection dominated. Matter falling in towards a star liberate $\sim GM/R$
of potential energy per unit mass. In a non-advection dominated flow most 
of this would be dissipated as heat in accretion disk, while in an advection
dominated flow this energy is carried in with the flow of matter. 
The heating rate is
\begin{equation}
 \dot{\epsilon}_h \sim \frac{R}{\lambda}\frac{GM\dot{M}}{R} 
 = 8\times 10^{51} M_{1.4}^{13/6}\dot{M}_{1.4} \,\,\,\,\, \mbox{erg s}^{-1}
\end{equation}
Finally we use the estimate of the heat capacity $C_v$ given in \cite{Watts:2001ej}
\begin{equation}
 C_v=1.6\times 10^{39}M_{1.4}^{1/3}T_{9} \,\,\,\,\, \mbox{erg K}^{-1}
\end{equation}

\section{Magnetic damping}

The crucial ingredient introduced in the previous Section is the magnetic
damping rate, which we have inserted in the evolution of the r-modes.
This specific modification of the r-mode equations is at the basis of the 
phenomenolgy we are going to discuss and was never considered in previous
calculations. The expression of the magnetic damping rate has been derived
in \cite{Rezzolla:2001di,Rezzolla:2001dh}, where it has been shown that
while the star remains in the instability region, the r-modes generate
a differential rotation which can greatly amplify a pre-existent magnetic
field. More specifically, if a poloidal magnetic field was originaly present,
a strong toroidal field is generated inside the star.
The energy of the modes is therefore transfered to the magnetic field and
the instability is damped.

The expression of the magnetic damping rate reads:
\begin{equation}
F_m\equiv \frac{1}{\tau_m}\simeq \frac{4(1-p)}{9\pi p \cdot (8.2\times 10^{-3})}
\frac{B_p^2 R \Lambda '\int^t_0 \alpha^2(t ')\Omega(t ') dt '}{M\Omega} ,
\end{equation}
where $p$ and $\Lambda '$ are dimensionless parameters of order unit.
The time integral over the r-mode amplitude $\alpha$ takes contribution
from the period during which the star is inside the instability region.
The crucial point is if the toroidal magnetic field generated 
during an instability phase can be stabilized or unwind on a
Alfv\'{e}n timescale.
\section{Results}
\begin{figure}
\begin{center}
\includegraphics[scale=0.7]{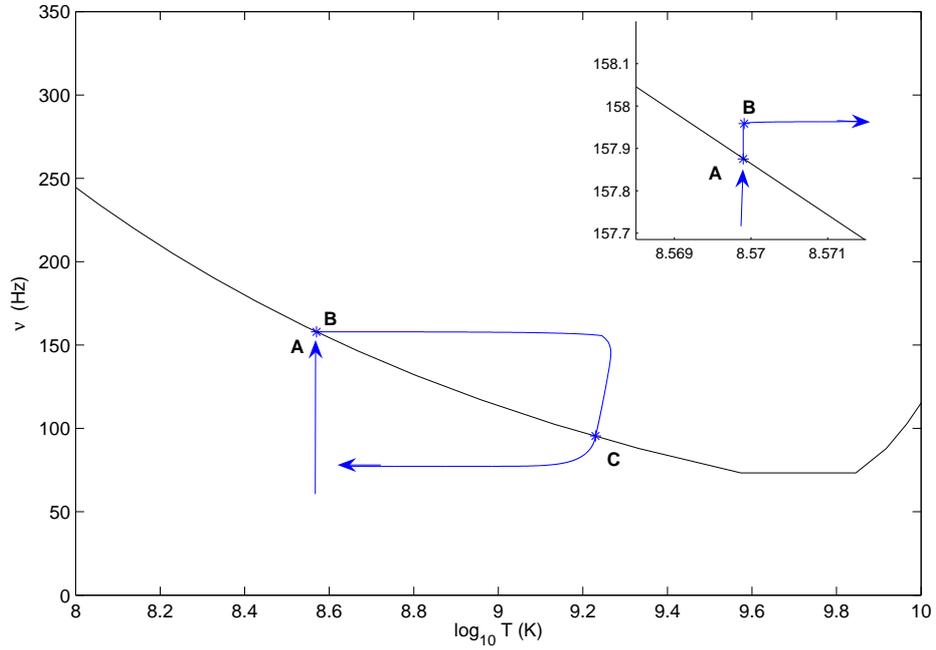}
\end{center}
\caption{\label{fig1} Path followed by the accreting neutron star, without 
toroidal magnetic field, in the Temperature vs Frequency plane. Here
$\dot{M}=10^{-8}\,\,\mbox{M}_{\odot}\,\,\mbox{yr}^{-1}$ and poloidal magnetic field $B_{pol}=10^8$~G.}
\end{figure}

\begin{figure}
\begin{center}
\includegraphics[scale=0.6]{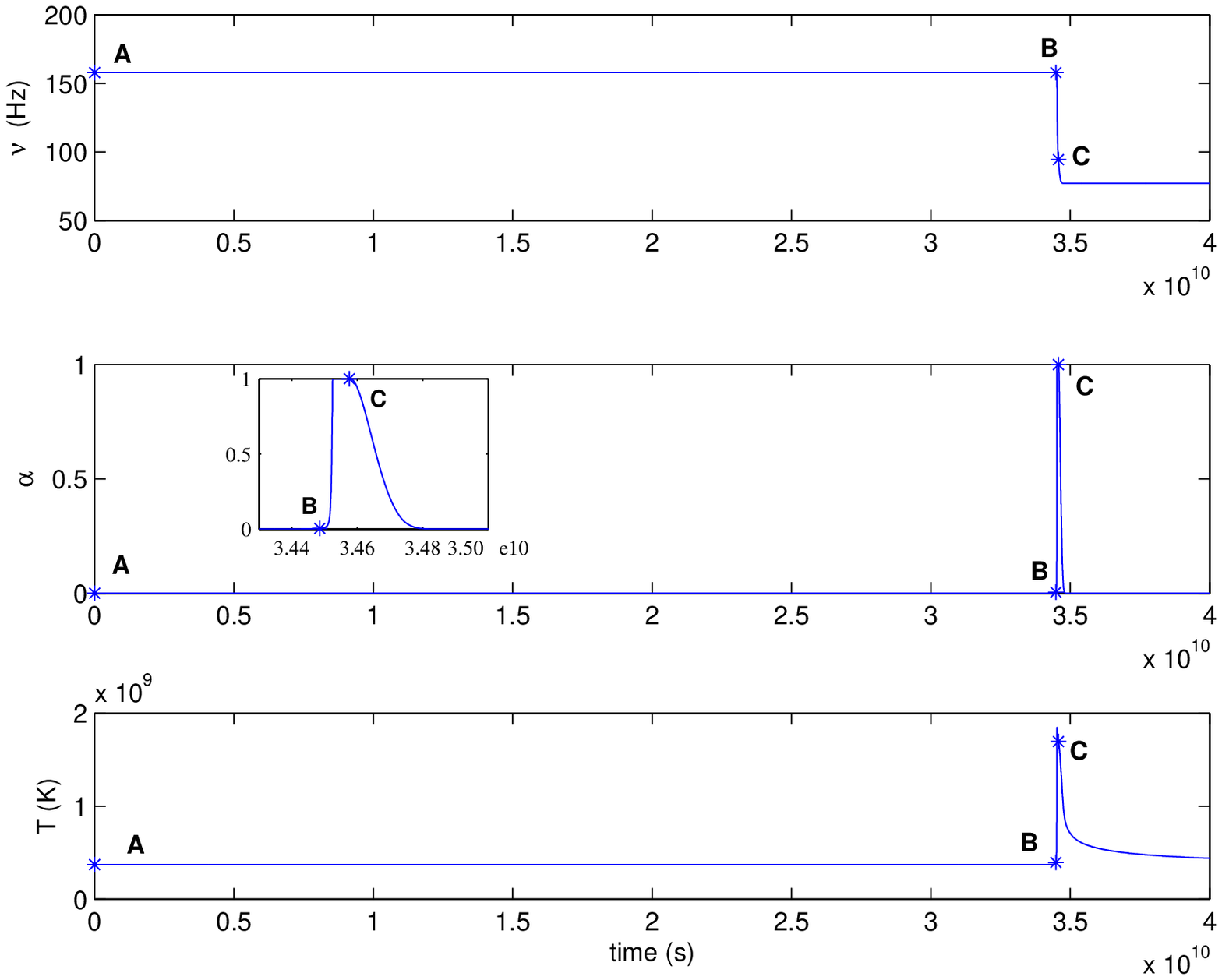}
\end{center}
\caption{\label{fig2}Temporal evolution, without toroidal magnetic field, of frequency, 
r-mode amplitude and temperature of the accreting neutron star.}
\end{figure}

\begin{figure}
\begin{center}
\includegraphics[scale=0.7]{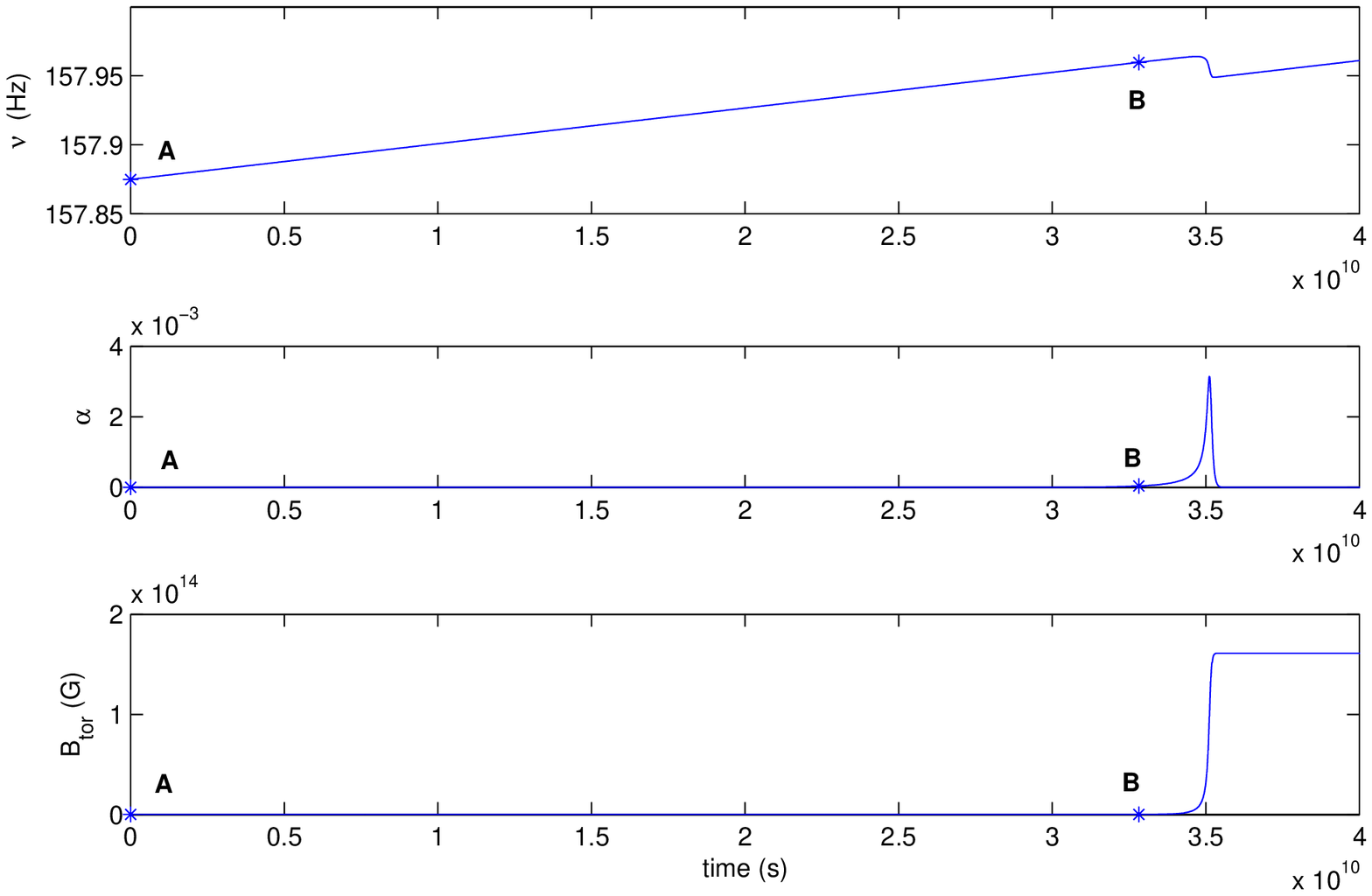}
\end{center}
\caption{\label{fig3}Temporal evolution, with toroidal magnetic field, of frequency, 
r-mode amplitude and temperature of the accreting neutron star.}
\end{figure}

We consider a scenario in which the mass accretion spin up a initially slowly
rotating neutron star and we investigate the case in which the star cannot find a state of
thermal equilibrium, then it enters a regime of thermogravitational
runaway within a few hundred years of crossing the r-mode stability 
curve. In Figure (\ref{fig1}) and Figure (\ref{fig2}) we show the evolution 
of temperature, spin frequency and r-mode amplitude obtained without magnetic field.
At the instant \textit{\textbf{A}} the accreting neutron star enters the r-modes 
instability window. Until the star is close to equilibrium the instability 
grows slowly and the star moves to higher temperatures mainly because 
of non thermal equilibrium. After the instant \textit{\textbf{B}} the 
r-mode instability grows exponentially due to the decrease of the shear viscosity 
with increasing temperature. As a consequence the r-modes amplitude rapidly 
reaches the saturation value and the viscosity heats significantly the star.
At this stage the star loses angular momentum by emission of gravitational waves
and within few hundred years goes out of the instability region (instant \textit{\textbf{C}}). \\
Taking into account the magnetic field, the evolutionary scenario for the star is quite different. 
After the instant \textit{\textbf{B}}, the secular effects are extremely large and the toroidal 
magnetic field is either produced or amplified by the wrapping of the poloidal magnetic field
produced by the (mostly) toroidal secular velocity field. 
Our results obtained solving Eqs.(\ref{eq5}) and reported in Figure (\ref{fig3}) show how
the new magnetic field so generated reaches a value $B_{tor}\sim 10^{14}$ 
G in about one hundred years and stabilizes the star when it is still
in the instability region. \\
The subsequent evolution of the system is difficult to control because it strongly
depends on the evolution of the magnetic field.
Analytical and numerical simulations show that strong purely toroidal fields are unstable.
We will limit ourselves to outline a possible scenario that we discuss in a forthcoming
paper. Following the results of some numerical simulations,
the magnetic field can evolve on an Alfv\'en timescale into a stable configuration
of a mixed poloidal-toroidal twisted-torus shape \cite{Braithwaite:2005md}.
In this scenario the newly formed magnetic field can provide a effective damping 
of the r-modes and rapidly rotating stars can be stabilized also in the absence
of the Ekman layer. \\
\newline
It is a pleasure to thank Giuseppe Pagliara and Andreas Reisenegger for many useful
and stimulating discussions.

\newpage
\section*{References}

\bibliography{art1}
\bibliographystyle{iopart-num}


\end{document}